\documentstyle[aps,psfig]{revtex}
\newcommand{\be}{\begin{equation}}
\newcommand{\ee}{\end{equation}}
\newcommand{\ben}{\begin{eqnarray}}
\newcommand{\een}{\end{eqnarray}}
\begin{document}

\twocolumn[\hsize\textwidth\columnwidth\hsize\csname 
@twocolumnfalse\endcsname

\title{Generation of Bright Soliton through the Interaction of Black Solitons} 
\author{L. Losano$^1$, B. Baseia$^2$, and D. Bazeia$^1$} 
\address{$^1$Departamento de F\'\i sica, Universidade Federal da Para\'\i ba,
58051-970 Jo\~ao Pessoa, Para\'\i ba, Brazil\\
$^2$Departamento de F\'\i sica, Universidade Federal de Goi\'as,
74001-970 Goi\^ania, Goi\'as, Brazil} 
\date{\today}

\maketitle

\begin{abstract}
We report on the possibility of having two black solitons interacting inside
a silica fiber that presents normal group-velocity dispersion, to generate
a pair of solitons, a vector soliton of the black-bright type. The model
obeys a pair of coupled nonlinear Schr\"odinger equations, that follows in
accordance with a Ginzburg-Landau equation describing the anisotropic $XY$
model. We solve the coupled equations using a trial-orbit method, which plays
a significant role when the Schr\"odinger equations are reduced to first order
differential equations.
\end{abstract}
\vskip2pc]

Solitary waves and solitons are relevant to a variety of nonlinear physical
processes. They have made perhaps the geatest experimental impact in the field
of nonlinear fiber optics, which are widely known to support solitons
of several different types \cite{agr95,hko95,klu98}.
For instance, in silica fibers there are optical solitons
of the bright and dark types. They appear because the Kerr
nonlinearity in silica is always described by a positive coefficient,
so that the sign of the group-velocity dispersion distinguishes
two different types of solitons in silica fibers: the bright soliton, for
negative sign of the group-velocity dispersion, and the dark soliton, for
positive group-velocity dispersion. These two types of optical solitons were
first reported in Ref.~{\cite{bright}}, in the case of bright solitons, and in
Ref.~{\cite{dark}}, in the case of dark solitons. They are solutions
of the cubic nonlinear Schr\"odinger equation, and they are characterized by
having distinct profile: the bright soliton has vanishing asymptotic behavior,
so they engender trivial topological behavior. On the other hand, dark
solitons are characterized by non vanishing asymptotic behavior, so the
presence of topological profile is one important feature of dark solitons.
The central characteristic of dark solitons is the presence of a dip in its
center, which distinguishes two possibilities: black solitons, in the case the
dip reaches the botton, that is, if the dip goes completely to zero, and gray
solitons, in the other cases.

Although both bright and dark solitons are widely known to be present in
fibers, they are not exclusive of fibers. They can also appear in other
systems, in particular in Bose-Einstein condensates (BEC), and in magnetic
materials. The old prediction that an ideal gas of identical bosons may
condensate into a macroscopic quantum state has recently been accomplished:
BECs of different atomic elements were first produced
in \cite{and95,dav95,bra97}, and are reviewed for instance in
Ref.~{\cite{par98}}. Soom after this achievment,
the presence of bright solitons in a BEC was first reported in
Ref.{\cite{jbc98}}, in the study of the dynamics of ultracold bosonic atoms
loaded in an optical lattice, that is, in a lattice induced by the
interference of an array of laser beams \cite{ol97}. The presence of dark
solitons was reported more recently in Refs.~{\cite{bbd99,den00}}.
The presence of bright or dark solitons in BEC is directly related to the
physics of attractive or repulsive atomic interaction, respectivelly,
in a mean-field description that follows in accordance with the
Gross-Pitaevskii equation \cite{dgp99}. In magnetic films one finds dark
solitons \cite{93,jap,prl} in the form of a microwave magnetic envelope (MME).
A good example of this was recently reported in Ref.~{\cite{ksp00}},
where one can find an interesting way of generating dark MME solitons,
which opens a new route for the investigation of dark solitons in magnetic
systems.

In silica fibers both black and bright solitons spring as solutions of the
cubic non-linear Schr\"odinger equation, with just a change in the
sign of the term that controls the group-velocity dispersion in that equation.
We illustrate the two possibilities considering the normalized equation that
describes the electric field envelope $U=U(x,z)$ inside the optical fiber
\be
\label{nlse1}
i\frac{\partial U_{\pm}}{\partial z}\pm\frac12
\frac{\partial^2U_{\pm}}{\partial x^2}+|U_{\pm}|^2
U_{\pm}=0
\ee
The variables $x$ and $z$ describe the reduced time and space coordinates
inside the fiber, respectively. This equation takes into accound the temporal
or group-velocity dispersion, the second term in the above Schr\"odinger
equation, and weak Kerr nonlinearity, the third term in Eq.~(\ref{nlse1}).
The two signs represent the two standard situations, that generate the black
and bright solitons.

The presence of solitons follows after writing
$U_{\pm}(x,t)=u_{\pm}(x)exp(i z)$. We get
\be
\frac12\frac{d^2u_{\pm}}{dx^2}=\pm\, u_{\pm}\mp u^3_{\pm}
\ee
These equations are solved by
\ben
u_+(x)&=&\pm{\rm sech}(x-{\bar x})
\\
u_-(x)&=&\pm\tanh(x-{\bar x})
\een
where ${\bar x}$ is an arbitrary point, standing for the center of the
soliton. Here $u_+(x)$ and $u_-(x)$ represent the fundamental solitons, the
bright and dark solitons, respectively. In fact, $u_-(x)$ is a black soliton
since the dip in $u^2_-(x)$ goes to zero, reaching the botton of the electric
field envelope.

We consider the possibility of describing a more general situation. We think
of using a single fiber, an optical medium with normal group-velocity
dispersion. However, we illuminate the fiber with two distinct laser beams,
having different amplitudes and opposite circular polarizations. The electric
fields are characterized by the envelopes $U(x,z)$ and $V(x,z)$, and the
system is described by the two normalized equations
\ben
\label{su}
i\frac{\partial U}{\partial z}-\frac12\frac{\partial^2 U}{\partial x^2}+
F(I)U&=&0
\\
\label{sv}
i\frac{\partial V}{\partial z}-\frac12\frac{\partial^2 V}{\partial x^2}+
G(I)V&=&0
\een
We use $I=|U|^2+|V|^2$ to describe the total intensity inside the fiber,
and now the envelopes of two interacting laser beams inside the optical
medium are characterized by the functions $F(I)$ and $G(I)$, which respond
to nonlinear Kerr interactions inside the silica fiber.

We describe the case of two different laser beams, that interact inside the
silica fiber with the envelope of the electric fields in the form
\ben
\label{hu}
U(x,z)&=&u(x)\;e^{i z}
\\
\label{hv}
V(x,z)&=&v(x)\;e^{irz}
\een
Here $r$ is real and positive, and parametrizes the relative propagation
constant of the vector constituents. We  substitute Eqs.~(\ref{hu}) and
(\ref{hv}) into Eqs.(\ref{su}) and (\ref{sv}) to obtain
\ben
\frac12\frac{d^2u}{dx^2}&=&-u+F_{r}(u,v)\;u 
\\
\frac12\frac{d^2v}{dx^2}&=&-r\;v+G_{r}(u,v)\;v
\een
We normalize the system in a way such that, in the absence of $v$ we get
$F_{r}(u,0)= u^2$, and in the absence of $u$ we get
$G_{r}(0,v)=r^2 v^2$. In the first case, for $v=0$ we get
\be
\frac12\frac{d^2u}{dx^2}=-u+u^3
\ee
whose solutions are
\be
\label{d1}
u_b(x)=\pm \tanh\left({x-{\bar x}}\right)
\ee
The second case is similar, leading to
\be
\label{d2}
v_b(x)=\pm \sqrt{\frac{1}{r}}
\tanh[\sqrt{r}\,(x-{\bar x})]
\ee
These solutions represent black soliton solutions. The system we are preparing
is such that in the reduced case of just one component, obtained after
turning off one of the two laser beams, the interactions between the optical
fiber and the beam give rise to a black soliton. Each one of the black
solitons has its own characteristics, which are naturally related to each
one of the two laser beams since we are not changing the optical medium,
the fiber. The individual black solitons are different, giving rise to an
asymmetrical arrangement that will allow interesting novelties. The idea is
similar to those in Refs.~{\cite{94,94a,97}}, and the results
remind us of other effects: in quantum optics, a superposition
of two (antisqueezed) number states may exhibit squeezing effect \cite{ss},
a property not shown by any isolated component; also, a superposition
of two (most classical) coherent states may result in a nonclassical
(Schr\"odinger's cat) state, as observed experimentally for light
fields \cite{cs} and for trapped ions \cite{ti}.

The main characteristics of the two black solitons are depicted in Fig.~1,
where we plot both the $u_b(x)$ and $v_b(x)$. We notice that the
amplitude ($a$) and width ($w$) of $u_b(x)$ are such that $a_u=w_u=1$,
but for $v_b(x)$ they become $a_v=w_v=\sqrt{1/r}$. Thus, in the following
we will consider the case $r\in(0,1)$.

\begin{figure}
\centerline{\psfig{figure=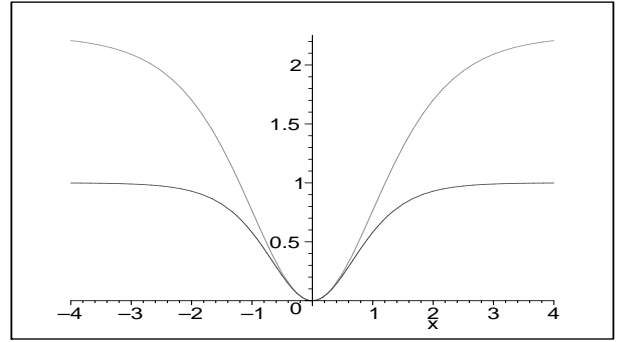,width=8.0cm,height=4.5cm}}
\vspace{0.3cm}
\caption{The two black solitons, $u^2_b(x)$ and $v^2_b(x)$, represented
by thicker and thinner lines, respectively. They appear inside the optical
fiber when one of the two laser beams is turned off. We use $r=4/9$.}
\end{figure}

The limit $r\to1$ describes the case of interactions between laser beams of
same amplitude but opposite circular polarizations in an isotropic Kerr
medium. This case is simpler, and can be described by a system of
symmetrically coupled nonlinear Schr\"odinger equations. For the pair of
symmetrically coupled equations, if we set $F_1(u,v)=G_1(u,v)=u^2+v^2$
we get the well known Manakov model \cite{man}. This model can also be used
to represent the isotropic $XY$ model \cite{wal97}, which obeys the
Ginzburg-Landau equation ${\partial A}/{\partial t}=
{\partial^2 A}/{\partial x^2}+A-|A|^2A$, where $A$ is complex. We use
$A=X+iY$ to show that for static configurations the above Ginzburg-Landau
equation leads to the symmetrically coupled nonlinear Schr\"odinger equations.
We use this as a guide toward the more general asymmetric model, and we
consider the anisotropic $XY$ model, which is described by
\be
\label{gla}
\frac{\partial A}{\partial t}=
\frac{\partial^2 A}{\partial x^2}+A-|A|^2 A +\alpha {\bar A}
\ee
where $\alpha$ is real. We use this model to investigate the case of two
laser beams, in a medium such that $F_{r}(u,v)=u^2+r(2r+1)v^2$ and
$G_{r}(u,v)=r^2v^2+r(2r+1)u^2$, that obey $F_{r}(u,0)=u^2$ and
$G_{r}(0,v)=r^2\,v^2$, as required to recuperate the case describing
a single laser beam, when the other beam is turned off. The form of the
interaction is dictated by the anisotropic $XY$ model, and by the two laser
beams that enter the fiber. 

The above considerations settle the problem. And now the key issue follows
after recognizing that the coupled Schr\"odinger equations can be seen as
equations of motion for static fields that appear in models of two real
scalar fields in bidimensional space-time \cite{raj82}. In the case under
examination the system of two fields is described by the potential
\be
V(u,v)=\frac12\left(-1+u^2+r\,v^2\right)^2+
\frac12\left(2r\, u\, v\right)^2
\ee
This potential has the general form \cite{95,96,00}
\be
\label{spot}
V(u,v)=\frac12\left(\frac{dW}{du}\right)^2+\frac12\left(\frac{dW}{dv}\right)^2
\ee
where $W=W(u,v)$ is given by
\be
W(u,v)=-u+\frac13 u^3+r u v^2
\ee
The pair of second order equations now become
\ben
\label{21}
\frac12\frac{d^2u}{dx^2}&=&-u+u^3+r(2r+1)uv^2
\\
\label{22}
\frac12\frac{d^2v}{dx^2}&=&-r v+r^2v^3+r(2r+1)u^2v 
\een
The main issue here is that the above system of nonlinear Schr\"odinger
equations can be solved exactly. There are two {\it trivial} solutions,
that appear when one turns off one of the two beams. They are the fundamental
black solitons $u_{1}(x)=u_b(x),\;v_{1}=0$ and
$u_{2}=0,\;v_{2}(x)=v_b(x)$, which reproduce the former solutions
(\ref{d1}) and (\ref{d2}). Furthermore, we also have the {\it non-trivial}
vector soliton, described by
\ben
\label{s21}
u_2(x)&=&\pm\tanh[{2r}(x-{\bar x})]
\\
\label{s22}
v_2(x)&=&\pm\sqrt{1/r-2\,}\;{\rm sech}[{2r}(x-{\bar x})]
\een
This result is very interesting: it shows that two black solitons in
interaction inside the silica fiber generate a pair of solitons, one of the
black type and the other of the bright type. In Fig.~2 we depict this
non-trivial vector soliton. It illustrate that the presence of bright solitons
is not a privilege of specific fibers, that have anomalous group-velocity
dispersion. We may say that the second laser beam sees the normal
group-velocity dispersion of the medium as an anomalous one, and this
appear through the presence of the first laser beam, which is responsible
for changing the group-velocity dispersion inside the fiber. This phenomenon
is new, and follows from the vector soliton (\ref{s21}) and (\ref{s22}). We
notice that the vector soliton only exists for $r\in(0,1/2)$, and the limit
$r\to1/2$ leads us back to the scalar soliton $u_b(x)$. This model has the
property that the limit $r\to1$ is trivial. This limit leads to the symmetric
situation, which can be investigated after rotating the the $(u,v)$ plane to
$(u_+,u_-)$, with $u_{\pm}=u\pm v$. These transformations factorate the
two-field system into two degenerate systems of a single field, that support
no vector field solutions of the black and bright type. This shows that our
model is asymmetric by construction, and the limit $r\to1$ leads to a trivial
symmetric model.

The presence of the nontrivial black-bright vector soliton is easier to see
if we integrate the coupled equations (\ref{21}) and (\ref{22}) once.
This gives the first order equations
\ben
\label{121}
\frac{du}{dx}&=&\mp1\pm u^2\pm r v^2
\\
\label{122}
\frac{dv}{dx}&=&\pm 2r u v
\een
The fact that the pair of second order
equations (\ref{21}) and (\ref{22}) can be integrated to give the above
pair of first order equations (\ref{121}) and (\ref{122}) is fundamental,
since it is easier to integrate. In Field Theory this identifies the
Bogomol'nyi-Prasad-Sommerfield bound \cite{bps}; the solutions
are stable and minimize the energy of the system \cite{95,96}.

\begin{figure}
\centerline{\psfig{figure=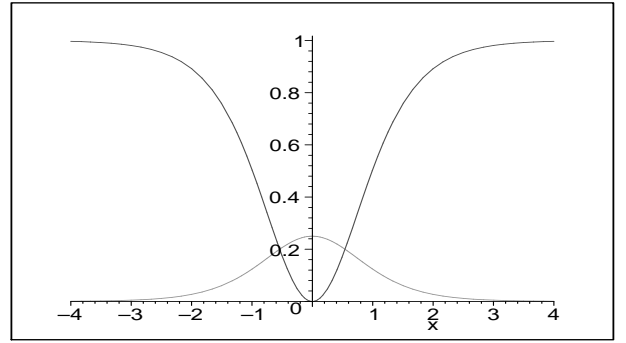,width=8.0cm,height=4.5cm}}
\vspace{0.3cm}
\caption{The black ($u^2_2(x)$, thicker line) and bright ($v^2_2(x)$,
thinner line) solitons that appear inside the optical fiber when both the
laser beams are turned on. We use $r=4/9$.}
\end{figure}

We solve the first order equations using the trial orbit method introduced in
Ref.~{\cite{raj79}}. We try the orbit $u^2+a v^2=1$, for $a$ being some
real constant. We
differentiate this expression and we use
the first order equations (\ref{121}) and (\ref{122}) to obtain
$u^2+r(1+2a)v^2=1$. This is compatible with the previous choice
if and only if $a=r/(1-2r)$. Thus the orbit is given by
\be
\label{o}
u^2+\frac{1}{\frac{1}{r}-2}\,v^2=1
\ee
It is an elliptical arc, that goes from $(u=\pm1,v=0)$ to
$(u=\mp1,v=0)$ when $x$ spans the entire real line.

We use the orbit (\ref{o}) to write the first order Eq.~(\ref{121}) in
the form ${du}/{dx}=\mp 2r(1-u^2)$. It can be integrated easily. The result
gives the solution (\ref{s21}). The other solution (\ref{s22}) is obtained
immediately from the orbit (\ref{o}) obeyed  by $u$ and $v$.

We have found the vector soliton as a solution
of the pair of first order equations (\ref{121}) and (\ref{122}). The
approach is new, and involves two steps: the first consists in transforming
the pair of second order Schr\"odinger equations (\ref{21}) and (\ref{22})
into a pair of first order differential equations; the second step deals
with the trial orbit method introduced in Ref.~{\cite{raj79}}. The trial
orbit method was introduced to help solving second order equations
of motion for relativistic systems of coupled scalar fields. In the
present context, the effectiveness of the trial orbit method is
direct, since we can use the first order equations themselves to check
if the orbit we trial is good, that is, if the trial orbit does not
contradict the first order equations that follow from the equations of
motions. This check is immediate, and provides a direct trial for the
eligibility of the orbit under investigation. This fact does not appear in
the original proposal in Ref.~{\cite{raj79}}, since there it deals with
second order differential equations.

We see that when the potential is written in terms of $W(u,v)$, in the
form (\ref{spot}), the two equations of motion are written as
\ben
\frac{d^2u}{dx^2}&=&W_u W_{uu}+W_v W_{vu}
\\
\frac{d^2v}{dx^2}&=&W_u W_{uv}+W_v W_{vv}
\een 
where $W_u=\partial W/\partial u$, etc. We associate to these equations
the pair of first order equations
\be
\frac{du}{dx}=W_u,\;\;\;\;\;\frac{dv}{dx}=W_v
\ee
We differentiate these first order equations to see that their solutions also
solve the equations of motion. Thus, we can concentrate on the simpler task
of solving first order equations to get solutions to the second order
equations. This is an important advantage, but it is restricted to work
when the composite vector soliton presents nontrivial topological feature,
as happens with dark solitons. This is not the case for bright solitons,
that engender trivial topological behavior. Thus we may wander if it is
possible to use two bright solitons to generate a composite, bright-black
vector soliton. Because the bright solitons are nontopological
they cannot appear as solutions of first order equations, as recently
shown in Ref.~{\cite{bms01}}.

We thank A. F. Lima and P. C. Oliveira for discussions, and CAPES,
CNPq, and PRONEX for partial support.

\end{document}